\documentclass[10pt,sigconf]{acmart}

\usepackage{multirow}
\usepackage{amssymb}
\usepackage{array}
\usepackage{booktabs}
\usepackage{tabularx}
\usepackage{wasysym}

\usepackage{hyperref}
\usepackage[T1]{fontenc}
\usepackage{caption}
\usepackage{subcaption}
\usepackage[whole]{bxcjkjatype}  

\usepackage{amsmath}
\usepackage{algorithm}
\usepackage{algcompatible}
\usepackage[noend]{algpseudocode}
\usepackage[normalem]{ulem}

\usepackage{footnote}
\makesavenoteenv{tabular}
\makesavenoteenv{table}

\algnewcommand{\LeftComment}[1]{\Statex \(\triangleright\) #1}

\renewcommand\footnotetextcopyrightpermission[1]{} 
\begin{document}


\author{Hiroaki Suzuki}
\affiliation{%
   \institution{Waseda University}
   \streetaddress{3-4-1 Okubo Shinjuku}
   \city{Tokyo} 
   \state{Japan} 
   \postcode{169–8555}
 }
\email{sz0213hk@nsl.cs.waseda.ac.jp}

\author{Daiki Chiba}
\orcid{0000-0002-7532-6633}
\affiliation{%
   \institution{NTT Secure Platform Laboratories}
   \streetaddress{3-9-11, Midori-cho, Musashino-shi}
   \city{Tokyo} 
   \state{Japan} 
   \postcode{180-8585}
 }
\email{daiki.chiba@ieee.org}

\author{Yoshiro Yoneya}
\affiliation{%
   \institution{Japan Registry Services}
   \streetaddress{--}
   \city{Tokyo} 
   \state{Japan} 
   \postcode{xxx-xxxx}
 }
\email{yone@jprs.co.jp}

\author{Tatsuya Mori}
\orcid{0000-0003-1583-4174}
\affiliation{%
   \institution{Waseda University/NICT/RIKEN AIP}
   \streetaddress{3-4-1 Okubo Shinjuku}
   \city{Tokyo} 
   \state{Japan} 
   \postcode{169–8555}
 }
\email{mori@nsl.cs.waseda.ac.jp}

\author{Shigeki Goto}
\affiliation{%
   \institution{Waseda University}
   \streetaddress{3-4-1 Okubo Shinjuku}
   \city{Tokyo} 
   \state{Japan} 
   \postcode{169–8555}
 }
\email{sgoto@waseda.jp}


\title[ShamFinder]{ShamFinder: An Automated Framework\\for Detecting IDN Homographs}
\titlenote{The paper is accepted at ACM IMC 2019 (\url{https://conferences.sigcomm.org/imc/2019/}). Please cite the IMC version.}

\begin{abstract}
The internationalized domain name (IDN) is a mechanism that enables us to use Unicode characters in domain names.
The set of Unicode characters contains several pairs of characters that are visually identical with each other; e.g., the Latin character `a' ({\tt U+0061}) and Cyrillic character `а' ({\tt U+0430}).
Visually identical characters such as these are generally known as {\it homoglyphs}.
{\it IDN homograph attacks}, which are widely known, abuse Unicode homoglyphs to create lookalike URLs.
Although the threat posed by IDN homograph attacks is not new, the recent rise of IDN adoption in both domain name registries and web browsers has resulted in the threat of these attacks becoming increasingly widespread, leading to large-scale phishing attacks such as those targeting cryptocurrency exchange companies.  
In this work, we developed a framework named ``{\it ShamFinder},'' which is an automated scheme to detect IDN homographs. 
Our key contribution is the automatic construction of a homoglyph database, which can be used for direct countermeasures against the attack and to inform users about the context of an IDN homograph. 
Using the {\it ShamFinder} framework, we perform a large-scale measurement study that aims to understand the IDN homographs that exist in the wild. 
On the basis of our approach, we provide insights into an effective countermeasure against the threats caused by the IDN homograph attack.

\end{abstract}


\maketitle

\section{Introduction}
\label{sec:intro}

Internationalized domain name (IDN) is a mechanism that allows us to use various non-English characters such as Arabic, Chinese, Cyrillic, Hangul, Hebrew, Hiragana, or Tamil. 
IDN was first proposed by  D\"{u}rst in 1996 as an Internet Draft (I-D)~\cite{idn}. Subsequently, a system known as Internationalizing Domain Names in Applications (IDNA) was adopted as an Internet standard~\cite{rfc3490}.
Currently, the IDNA system is widely deployed in various domains including hundreds of top-level domains (TLDs).
In addition, the majority of modern web browsers are capable of accommodating IDNs.

Character sets permitted to be used as IDNs contain several pairs of characters that are visually similar with each other. These characters are known as {\it homoglyphs}.
The existence of homoglyphs enables an attacker to create a spoofing domain name. 
For instance, by using a Unicode character `\'e', which is a Latin lowercase letter e with an acute accent (U+00E9), an attacker can create a spoofing domain name, ``fac\'ebook.com,'' which is visually similar to the original domain name  ``facebook.com.'' 
The domain spoofing attack exploiting Unicode homoglyphs is known as ``IDN homograph attack'' and has been used for malicious purposes such as phishing attacks. 
IDN homograph attacks are not a new problem. In 2002, Gabrilovich and  Gontmakher~\cite{DBLP:journals/cacm/GabrilovichG02} demonstrated that they successfully registered an IDN homograph using the two Russian letters `с' and `о'.

As the adoption rate of IDN was not high in the past, an IDN homograph has been recognized as a proof-of-concept attack. However, the recent rise in the number of IDN registrations and the adoption of an IDN in many TLDs together with the adoption of IDNs in modern browsers have resulted in the threat of IDN becoming realistic and has attracted interest from researchers~\cite{DBLP:conf/dsn/LiuLLLDHZ18,zheng-blog2017} as well as from attackers. 
In May 2018, Binance, which is a cryptocurrency exchange company, reported that their primary domain name \url{binance.com} was the victim of an IDN homograph attack~\cite{binance2018}. 
We note that as this incident implied, the targets of IDN homograph attacks are not only browsers, but also email clients, where a victim could click a malicious URL composed of an IDN homograph.

A straightforward and effective countermeasure against the threat of an IDN homograph attack is to identify possible IDN homographs. 
The key technical challenge here is to automate the process of detecting homoglyphs that could be abused for creating an IDN homograph. 
As of May 2019, of the 137,928 characters included in Unicode~12.0.0~\cite{unicode12}, 123,006 characters can be used for IDN, following the specification of {\it IDNA2008}~\cite{faltstrom-unicode12}. 
Furthermore, the number of IDNs registered has continued to increase.  
According to the IDN World Report~\cite{IDN-world18}, 
the estimated number of IDNs registered was 2.0 million in 2009 and this number increased to 7.5 million IDNs in December 2017.

In this work, we developed a generic framework named ``{\it ShamFinder},'' which aims to identify IDN homographs in a scalable manner.
The key technical contribution of {\it ShamFinder} is to build a new homoglyph database named {\it SimChar}, which can be maintained without requiring time-consuming manual effort.
Unlike previous approaches for detecting IDN homographs~\cite{DBLP:conf/dsn/LiuLLLDHZ18,sawabe18}, the notable advantage of {\it ShamFinder} is that it can pinpoint the differential characters; thus, it can be used for direct countermeasures such as building a blacklist of the confusable characters or highlighting the anomalous characters to inform the user of the potential risk of an IDN homograph attack.
We note that our homoglyph database covers a wide range of homoglyphs that have not been listed in the existing database maintained by the Unicode consortium~\cite{confusables}.

Using the {\it ShamFinder} framework, we attempt to understand the IDN homographs registered in the wild. 
In our study, we investigated the way in which the registered IDN homographs are abused by collecting IDNs from the world's most popular TLD, {\tt .com}.
In addition, using {\em ShamFinder} as a building block, we discuss a proof-of-concept system that aims to mitigate the threats posed by an IDN homograph attack.

The main contributions of this work are summarized as follows:
\begin{itemize}
\item We developed a framework named {\it ShamFinder}, which aims to identify IDN homographs in an automated manner.   
\item We built a new homoglyph database named {\it SimChar}, which can be automatically updated and can be used for other security applications such as detecting plagiarism that exploits homoglyphs.
\item Using the {\it ShamFinder} framework, we performed a large-scale measurement study on how IDNs are used or abused in the wild. The measurement study demonstrated that our framework efficiently extracted IDN homographs, which contained malicious ones.
\item Based on the {\it ShamFinder} framework, we propose a practical countermeasure against the generic threat of IDN homograph attacks.
\end{itemize}

The remainder of the paper is organized as follows:
Section~\ref{sec:background} presents an overview of IDN and IDN homograph attacks. 
In Section~\ref{sec:framework}, we introduce the {\it ShamFinder} framework. 
Section~\ref{sec:evaluation} contains an evaluation of the performance of the {\it ShamFinder} framework from the viewpoints of human perception and computational costs.
In Sections~\ref{sec:data} and~\ref{sec:measurement}, we present our data sources and findings derived from the large-scale measurement of IDN in the wild, using the {\it ShamFinder} framework. 
Section~\ref{sec:discussion} discusses the limitations of our work and effective countermeasures against the threats posed by IDN homograph attacks. 
In Section~\ref{sec:related}, we review related work in comparison with ours. 
We conclude our work in Section~\ref{sec:conclusion}.
\section{Background}
\label{sec:background}

This section first presents an overview of IDNs. 
We then provide an overview of IDN homograph attacks and recent studies on the threats posed by these attacks.

\subsection{IDN and Permitted Unicode Characters}

Since the initial proposal of IDN in 1996, its protocol specification has been standardized.
In 2003, the Internet Corporation for Assigned Names and Numbers (ICANN) and top IDN registries such as {\tt .cn}, {\tt .info}, {\tt .jp}, {\tt .org}, and {\tt .tw} have published a guideline for the implementation/operation of IDN~\cite{iana-guideline}． 
The guideline requires TLD registries to employ an ``inclusion-based'' approach, i.e., in each TLD, only code points that are permitted by the TLD can be used for IDN. 
Each TLD employs language-specific registration and administration rules, which are publicly available as IDN tables~\cite{idn-table}.
The tables are maintained by the Internet Assigned Number Authority (IANA).

This restriction introduced by the inclusion-based approach is expected to thwart the threats of IDN homograph attacks because the set of characters that can be used for IDN are limited with the tables.
For instance, the JP domain, which is the country code top-level domain (ccTLD) for Japan, limits the permitted character sets for IDN to LDH, which consists of case-insensitive English letters, digits, and hyphens (Letter-Digit-Hypen),
Hiragana, Katakana, and a subset of CJK unified ideographs  (character set used in Chinese, Japanese, and Korean). 
Therefore, it is not possible to register Latin-based IDN homographs with names such as ``\'acm.jp'' because the permitted characters for IDN of the JP domain do not contain a homoglyph of LDH.

However, as we shall present later, among the characters permitted for each TLD such as .com, there are many homoglyphs, indicating that an attacker can leverage such homoglyphs to execute an IDN homograph attack.
We note that an attacker can create an IDN homograph of a non-Latin IDN homograph. 
One of the key contributions of our work is to automatically build a comprehensive list of homoglyphs, which could be potentially abused for IDN homograph attack.

Although the IDN extension allows us to use non-Latin characters for domain names, we need to use LDH at the protocol level for backward compatibility reasons. 
Therefore, we need a mechanism that transcodes a domain name consisting of Unicode characters into one with LDH characters.
In this regard, Punycode is a character encoding scheme for transcoding a Unicode string to a string with LDH. The specification of Punycode is defined in RFC 3492~\cite{rfc3492}.
When using a string transcoded by Punycode for IDN, we add the prefix ``\verb|xn--|'' to the beginning of the transcoded string. 
For instance, the string ``阿里巴巴'' is represented as ``tsta8290bfzd'' by the Punycode transcoding, and the corresponding IDN is ``\verb|xn--tsta8290bfzd.com|''.

Finally, we note that each web browser implements the processing of IDN in a different way~\cite{firefox-idn, chrome-idn}.
As we explain below, the way IDN is displayed in the address bar could increase or decrease the threat of an IDN homograph attack.
Thus, the implementation largely affects the way users react to the IDN homograph presented in a browser. 
In Section~\ref{sec:discussion}, we discuss a proof-of-concept implementation of IDN processing on a browser to enable users to become knowledgeable of the existence of a possible IDN homograph attack without sacrificing the usability of IDN for them. 

\subsection{IDN Homograph Attack}
As mentioned in the previous section, the history of IDN homograph attacks can be traced back to the early 2000s.
As Gabrilovich and Gontmakher~\cite{DBLP:journals/cacm/GabrilovichG02} reported in 2002, numerous English domain names can be {\it homographed} by leveraging non-Latin letters. 

Despite the fact that threats of IDN homograph attacks were pointed out earlier, effective and usable countermeasures against these threats have not been developed. 
We conjecture that the reason behind abandoning the threats is that IDN has not been widely deployed in the world and there have been few web clients that can correctly process IDN.
However, the situation has changed because popular web browsers today have developed the ability to handle IDNs. 
In addition, according to the IDN World Report~\cite{IDN-world18}, 7.5 million of IDNs have been registered by December 2017.
These observations imply that the threat of IDN homograph attacks have become real.  
In fact, as mentioned in Section~\ref{sec:intro}, the cryptocurrency exchange company Binance was the victim of an IDN homograph attack.

As countermeasures against IDN homograph attacks, many browser vendors have updated the implementation of displaying IDN in the address bar after the threat of an IDN homolog attack was widely publicized by a blog post on the web~\cite{zheng-blog2017} in April 2017.
Specifically, Firefox and Chrome have changed their implementations as follows: when characters originating from multiple scripts (character sets) are mixed in a character string constituting an IDN, the IDN is displayed in the form of Punycode instead of Unicode~\cite{firefox-idn, chrome-idn}.
For instance, if a Latin-script-based domain name comprises non-English scripts such as Latin scripts, Cyrillic scripts, or Greek scripts, the domain name is displayed in the form of Punycode; i.e., for ``facébook'', its Punycode, \verb|xn--facbook-dya| is displayed in the address bar.

Although this update can be expected to mitigate the threats of IDN homographs to some extent, it is likely to impair the usability because Punycode is not a human-friendly representation. As the human-readable domain name provides hints as to the authenticity of the website, masking the original domain name may leave users less knowledgeable.
Although the aforementioned countermeasure by the browsers becomes a temporary countermeasure against IDN homograph attacks, if it is compulsorily displayed in Punycode, it is problematic in that it becomes difficult to understand the cause of the threat. That is, because the user does not notice that the domain name entered in the browser is a homograph attack, the user risks visiting the site with the same domain name again.

We also note that, in the above implementations, even in the case of an IDN composed of multiple scripts, if the domain name comprises both Latin script and a CJK ideograph, it will be displayed with Unicode. 
Furthermore, an attacker can create an IDN by not only combining Latin script, Cyrillic script, or Greek script but also by combining characters belonging to the set of CJK ideographs. 
We refer to such a homograph as a non-Latin homograph. 
For instance, the string ``工業大学'' (meaning an institute of technology in English) has the homograph, ``エ業大学'', where `工' is 
a CJK Unified Ideograph (U+5DE5) and `エ' is a Katakana Letter (U+30A8). 
Current web browsers do not have a way to identify non-Latin IDN homographs such as this.
\section{ShamFinder Framework}
\label{sec:framework}

In this section, we first provide a high-level overview of the {\it ShamFinder} framework.
Next, we present several Unicode character sets and those used for IDN. We note that precise understanding of these character sets is essential in extracting Unicode homoglyphs that could be abused for an IDN homograph attack.
We then describe the approach we followed to build the homoglyph database, which plays a key role in the {\it ShamFinder} framework. 
Finally, we describe the characteristics of the homoglyph database.

\subsection{High-level Overview}

\begin{figure}[tbp]
    \begin{center}
    \includegraphics[width=\linewidth]{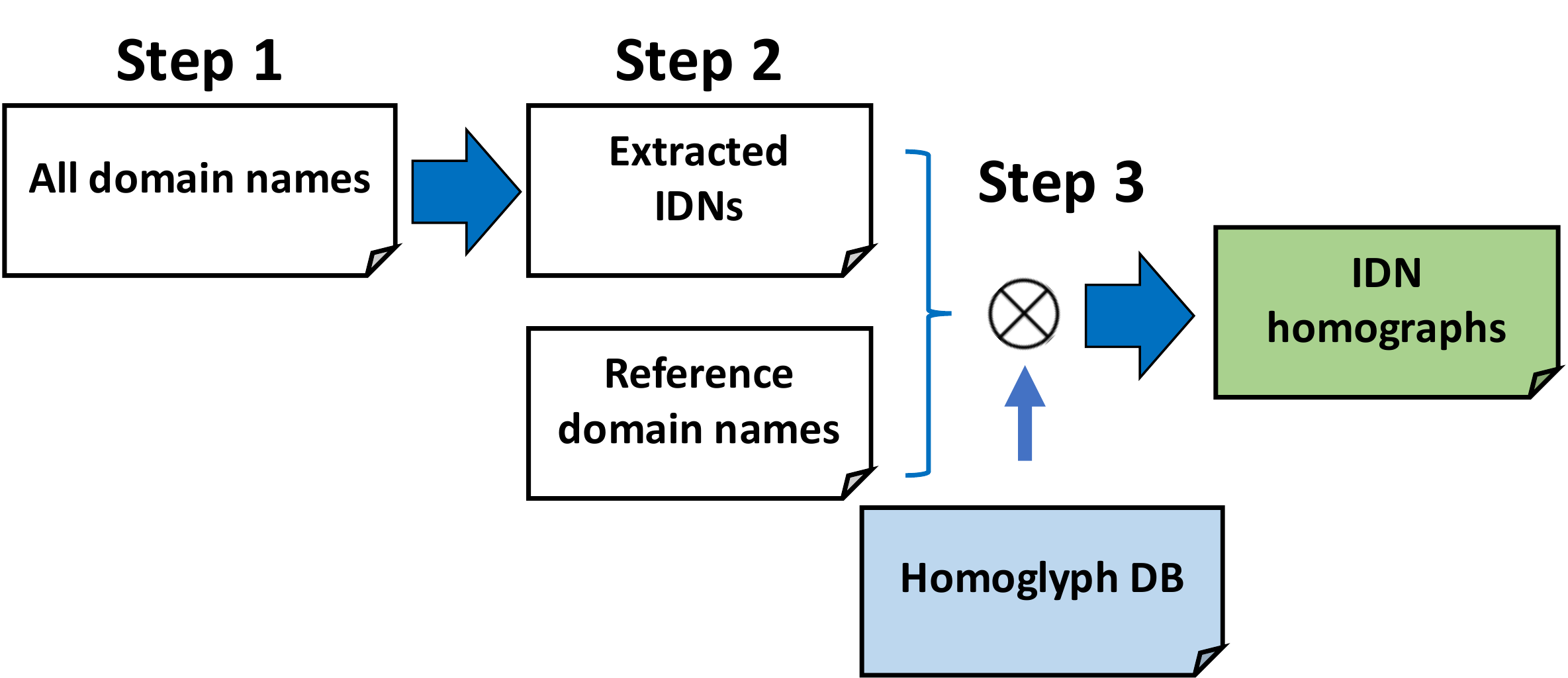}
    \caption{Overview of the ShamFinder Framework.}
    \label{fig:overview}
    \end{center}
\end{figure}

Figure~\ref{fig:overview} presents a high-level overview of the {\it ShamFinder} framework. 

\noindent{\bf Step 1}: First, we collect registered/active domain names for each TLD. 
To this end, we can either make use of the DNS zone file for each TLD or publicly available/commercial domain name lists such as~\cite{domainlists-io}. We introduce the datasets we used for our analysis in Section~\ref{sec:data}.

\noindent{\bf Step 2}: Next, we extract IDNs from the collected domain names by searching for those starting with the prefix ``\verb|xn--|.''

\noindent{\bf Step 3}: To find IDN homographs, we leverage a list of popular domain names as reference. As representative reference we can leverage a website ranking lists~\cite{DBLP:conf/imc/ScheitleHGJZSV18,DBLP:conf/pam/RweyemamuLWRK19} such as Alexa Top Sites~\cite{alexa} or Majestic Million~\cite{majestic}.
Next, we leverage the database of homoglyphs to identify potential IDN homographs; as we show in the next subsection, our contribution is to present a way of automatically building such a database.

\begin{figure}[tbp]
    \begin{center}
    \includegraphics[width=\linewidth]{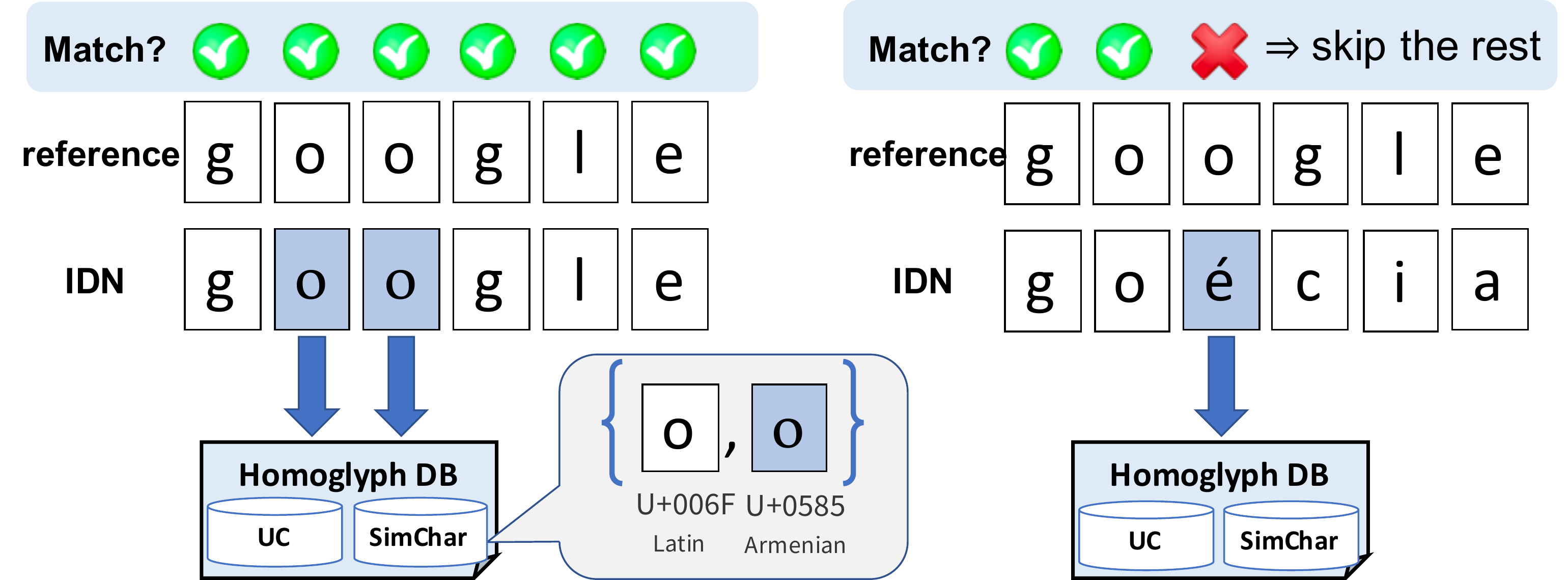}
    \caption{IDN homograph detection scheme for a given TLD. Left: a domain name is detected as an IDN homograph. Right: a domain name is not detected as an IDN homograph. In this example, the TLD part has been removed. }
    \label{fig:detect}
    \end{center}
\end{figure}

Figure~\ref{fig:detect} and Algorithm~\ref{alg:detection} show the IDN detection scheme. 
We check the length (number of characters) of each domain name listed in the reference domain names list and extract the IDNs with the same number of characters. 
For each pair consisting of a reference domain name and sampled IDN, we check their letters one by one to determine whether they correspond. If two corresponding letters match each other, we proceed to the next pair of letters. If the letters do not match, we check whether the pair is listed in the homoglyph database, which we present in the next subsection. If they are listed, we proceed to the next pair of letters and repeat the same process. If we find letters that do not match, we conclude that the IDN is not an IDN homograph of the reference domain name. 
The computational complexity of the algorithm is $|N||M||L|$ where $|N|$, $|M|$, and $L$ are the number of reference websites, number of IDNs, and number of characters contained in a domain name, respectively.
Although this is a na\"{\i}ve approach, the actual calculation cost has been reduced by restricting the computation of matching to those pairs of strings with the same length. The evaluation of the time needed for the computation appears in Section~\ref{sec:evaluation}.

\begin{algorithm}[tbp]
  \caption{Homograph detection algorithm}   
  \label{alg:detection}
  \begin{algorithmic}
  \footnotesize
\State {$\Omega \Leftarrow$ A set of IDNs for a given TLD (TLD part removed)}
\State {$R \Leftarrow$ A set of reference domain names for a given TLD (TLD part removed)}
\LeftComment{$len(x)$ returns the length of a string $x$.}  　
\LeftComment{$x[i]$ represents the $i$-th character of string $x$.}

\For{$r \in R$}
  　\State Let $\omega \subset \Omega$ be a set s.t. $len(r) = len(x)$ $\forall x \in \omega$
    \For{$x \in \omega$}
      \For{$i=1; i\leq len(x); i\to i+1 $}
        \If{$r[i] = x[i]$}
          next
        \ElsIf{$r[i]$ and $x[i]$ are listed as a pair in the SimChar database}
          next
        \Else
          \State Mark $x$ as not being a homograph of $r$ and skip the loops for the next $x$
        \EndIf
      \EndFor
      \State Mark $x$ as a homograph of $r$
    \EndFor  　
\EndFor
\end{algorithmic}
\end{algorithm}

\subsection{Unicode Characters Sets and IDN}
\label{sec:simchar}

\begin{figure}[tbp]
    \begin{center}
    \includegraphics[width=0.8\linewidth]{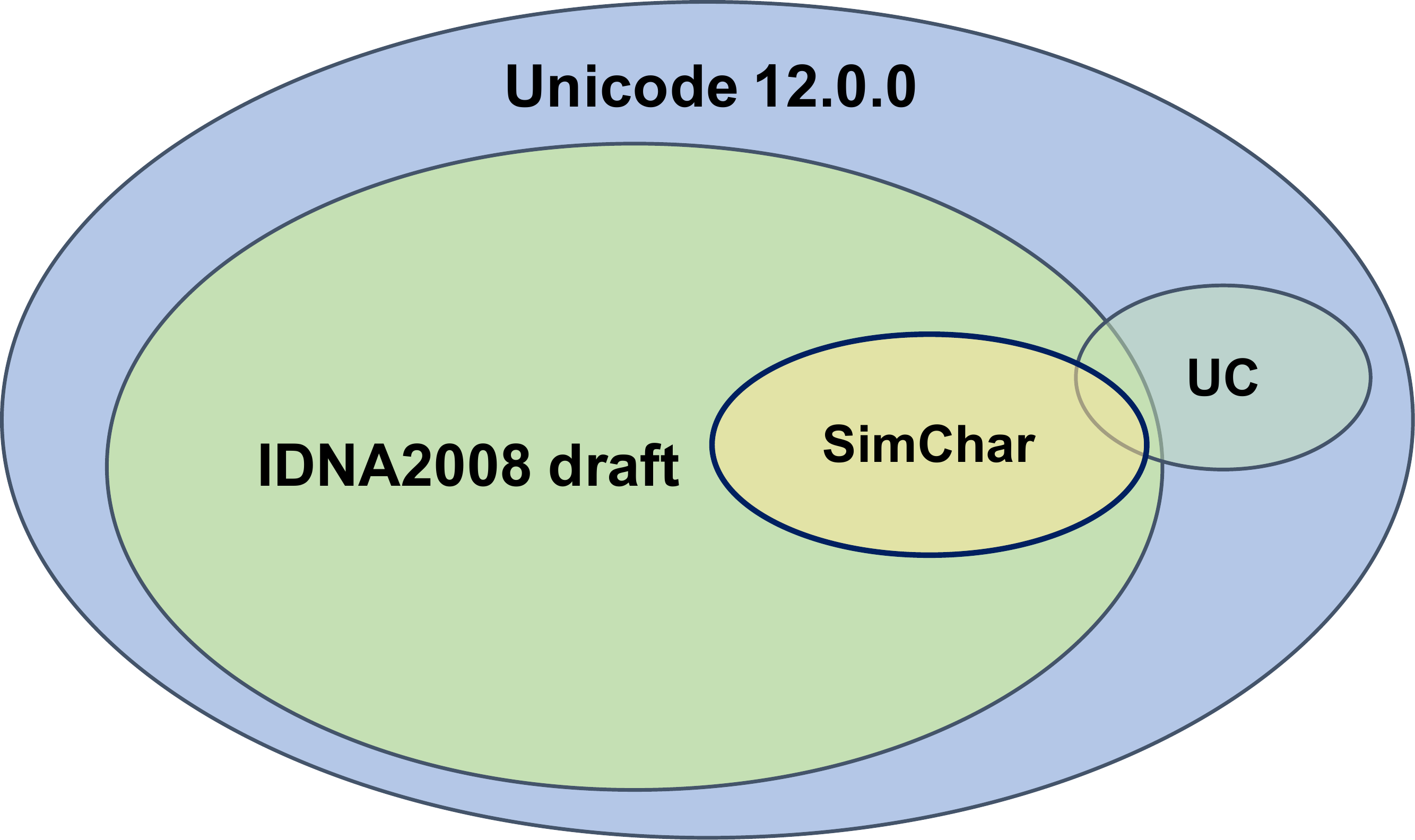}
    \caption{Contamination and overlap of character sets. UC denotes the recommended confusable mapping for IDN, ``confusables.txt.''}
    \label{fig:charsets}
    \end{center}
\end{figure}

Our primary goal is to compile a database that lists pairs of visually identical Unicode characters (homoglyphs) that are permitted to be used for IDN. 
We explain how we compile this database by beginning with a description of several Unicode character sets.
Figure~\ref{fig:charsets} summarizes the contamination and overlap of the Unicode character sets. 
The root set is the characters contained in Unicode 12.0.0~\cite{unicode12}. 
The set contains a total of {137,928} characters, covering {150} scripts, including modern/historic characters, signs, and symbols such as Emoticons.
Of the character sets defined in Unicode 12.0, the latest set of characters permitted for the use in IDN is defined in the Internet draft, named draft-faltstrom-unicode12-00 (``IDNA2008 and Unicode 12.0.0'')~\cite{faltstrom-unicode12}.
The number of Unicode characters contained in the IDNA2008 draft is {123,006}; these characters are listed in the section, ``Code points in Unicode Character Database (UCD),'' of the draft with the property of ``PROTOCOL VALID (PVALID),'' which indicates that the code points with the property value are permitted for general use in IDNs~\cite{rfc5892}.

\begin{table}[tbp]
\caption{Number of characters contained in each set shown in Figure~\ref{fig:charsets}; IDNA2008 draft (IDNA), Unicode confusables.txt (UC), and SimChar.}
\label{tab:charset_breakdown1}
\centering
  \begin{tabular}{l|r|r} \hline
  Sets &  \# characters & \# homoglyph pairs\\\hline\hline
  IDNA & 123,006 & n/a \\
  \hline
  {UC}  & 9,605 & 6,296 \\ 
  {UC} $\cap$ IDNA & 980    & 627 \\
  {\it SimChar}       & 12,686 & 13,208 \\
  {\it SimChar} $\cap$ {UC} & 233  & 127 \\
  {\it SimChar} $\cup$ (UC $\cap$ IDNA) & 13,210 & 13,708 \\
  \hline
  \end{tabular}
\end{table}

In the document named Unicode Technical Standard \#39 (UNICODE SECURITY MECHANISMS), a database named ``confusables.txt'' is provided. 
This text file compiles the confusable mapping for IDN. 
In this work, we refer to the database as {\it UC} for brevity. 
The {UC} database lists visually {\it confusable} characters and provides a mapping for visual confusables for use in detecting security problems such as an IDN homograph attack.
Although UC covers a wide range of homoglyphs that could be abused for IDN homograph attacks, our empirical observations revealed that a non-negligible number of homoglyphs are {\it not} contained in UC as shown in Table~\ref{tab:charset_breakdown1}.
This observation motivated us to build a new homoglyph database, {\it SimChar}, which is described in the next subsection. 
We note that although UC has been manually maintained, we can build {\it SimChar} in an automated way, implying that it can discover new homoglyphs from newly registered Unicode characters in future.  
Furthermore, as explained in Section~\ref{sec:evaluation}, homoglyphs contained in {\it SimChar} are more confusing than those contained in {UC}.

We note that {UC} covers several characters that are {\it not} contained in the IDNA2008 draft.  
Of the characters defined in the IDNA2008 draft, {980} characters are listed in {UC}; i.e., these {980} characters are potentially abused for IDN homograph attacks.
Our contribution is to build a complementary database named {\it SimChar}, which is compiled of a set of characters that have at least one homoglyph character from the IDNA2008 draft character set.
The new set has {13,210} characters that are also included in {UC}. It adds {3,605} characters that have not been listed in {UC}. Moreover, as seen in Table~\ref{tab:latinhomoglyphs}, {\it SimChar} adds {316} homoglyphs of Basic Latin characters that are {\it not} listed in {UC}. 
Table~\ref{tab:charset_breakdown1} summarizes the number of characters contained in the character sets shown in Figure~\ref{fig:charsets}.
We note that the {\it ShamFinder} framework makes use of the union of two sets {UC} and {\it SimChar} to find IDN homographs.
We also note that a character could be the homoglyph of several other characters. We count such pairs as ``Homoglyph pairs.'' Homoglyphs contained in {\it SimChar} are built from a set of characters contained in IDNA.
We notice that although the number of characters contained in {UC} is roughly 10K, if we consider the number of IDNA-permitted characters, the size becomes much smaller by a factor of 10. The details of {\it SimChar} will be shown later.

\begin{figure}[tbp]
    \centering 
    \includegraphics[width=0.8\linewidth]{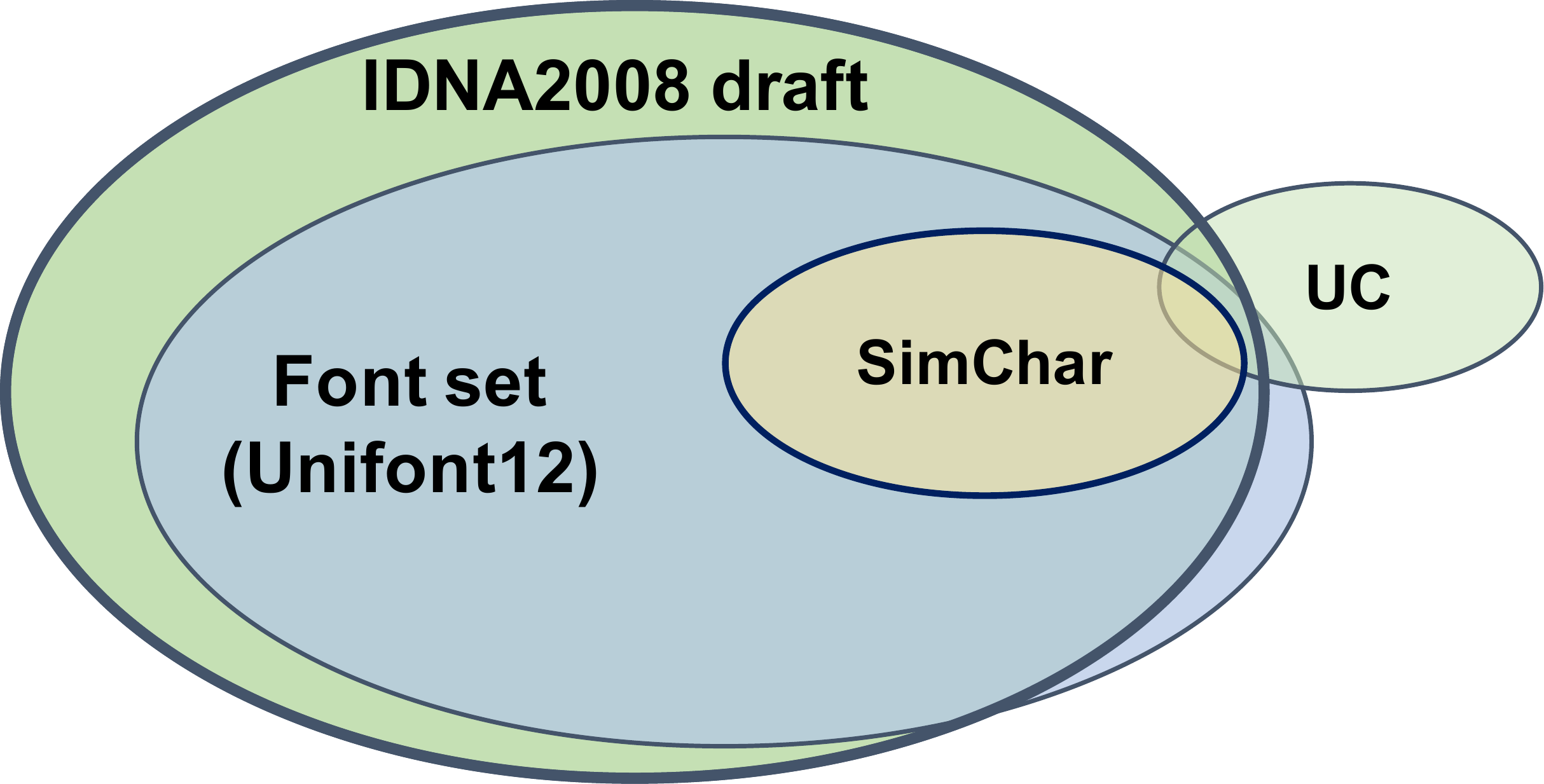}
    \caption{Contamination and overlap of character sets.}
    \label{fig:charsets2}
\end{figure}

\subsection{Building Homoglyph Database}
\label{sec:homoglyphdb}
As shown in Fig.~\ref{fig:detect}, we use UC and {\it SimChar} as the components of the homoglyph database we used to detect IDN homographs.
The key idea of {\it SimChar} is to extract homoglyphs by computing the similarity between the glyphs of corresponding characters. 
We first need to represent each code point as a visual image (glyph). 
To this end, we can make use of various Unicode fonts such as those listed in~\cite{unicode-fonts-wikipedia}. 
In this work, we adopt GNU Unifont Glyphs~\cite{unifont}, which covers the entire collection of characters contained in the Unicode Basic Multilingual Plane (BMP) as well as several other characters of the Supplemental Multilingual Plane (SMP).
Although BMP contains characters for almost all modern languages and a large number of symbols, SMP contains historic characters and signs as well as the symbols used in various fields such as Emoticons.
Even though the choice of a font may affect the detected homoglyphs, the following procedure can easily be extended to other font sets. We aim to evaluate other fonts in future work.

Figure~\ref{fig:charsets2} presents the relationship between the character sets.
Of the characters contained in the IDNA2008 draft, the latest version of Unifont (Unifont12 for short) covers {52,457} characters. 
Several IDN-permitted characters are {\it not} covered by Unifont12. However, as Unifont provides much larger coverage than other proprietary Unicode fonts such as Microsoft JhengHei, we deem the choice to be reasonable. 
In fact, of the 2,990 IDN-permitted characters in {UC}, 2,877 characters are covered by Unifont12.
Table~\ref{tab:charset_breakdown2} summarizes the number of characters contained in the character sets shown in Figure~\ref{fig:charsets2}.
In the following, we denote UC and {\it SimChar} as those with the union sets of Unifont12 for brevity.

\begin{table}[tbp]
\caption{Number of characters contained in the Unifont12 font set (Figure~\ref{fig:charsets2}).}
\label{tab:charset_breakdown2}
\centering
\vspace{-2mm}
  \begin{tabular}{l|r|r} \hline
  Sets &  \# Chars & \# Pairs\\\hline\hline
  IDNA $\cap$ Unifont12 & 52,457 & n/a\\
  \hline
  UC $\cap$ Unifont12 & 5,080 & 3,696\\
  {\it SimChar} $\cap$ Unifont12\footnote{Note that {\it SimChar} is composed using the union set of IDNA and Unifont12. Therefore, $|${\it SimChar} $| = |$ {\it SimChar} $\cap \mbox{Unifont12}|$.} & 12,686 & 13,208\\
  \hline
  \end{tabular} 
\end{table}

Next, we attempt to identify homoglyphs by testing their similarity as images. 
The structural similarity index measure (SSIM) is a widely used metric to quantify the degradation of image quality caused by processing methods such as data compression or by losses in data transmission~\cite{DBLP:journals/tip/WangBSS04, DBLP:conf/icpr/HoreZ10}. 
Thus, it can also quantify the similarity between a pair of images. 
However, because our objective is not assessing the perceptual metric that quantifies image quality degradation, we directly count the number of different pixels between two images. 
Let $I(i,j)\in\{0,1\}~(0\leq i \leq N-1, 0\leq j \leq N-1)$ be a square image having $N\times N$ pixels, where each pixel is represented as a binary digit.  
Our metric, $\Delta$ is computed as 
\begin{equation*}
    \Delta = \sum_{i=0}^{N-1}\sum_{j=0}^{N-1}\left|I_1(i,j)-I_2(i,j)\right|.
\end{equation*}
When $\Delta=0$, it indicates that two images are completely identical.

We note that $\Delta$ can be associated with the peak signal-to-noise ratio (PSNR), which is another widely used metric aimed at quantifying the reproducibility of images~\cite{DBLP:journals/tip/WangBSS04, DBLP:conf/icpr/HoreZ10}. 
In our model, $I(i,j)$ is represented as a binary bit. Therefore, the mean square error (MSE) is computed as 
\begin{equation*}
    MSE = \frac{1}{N^2}\sum_{i=0}^{N-1}\sum_{j=0}^{N-1}\left\{I_1(i,j)-I_2(i,j)\right\}^2 = \frac{\Delta}{N^2} .
\end{equation*}
Using the MSE, the PSNR is computed as 
\begin{equation*}
    PSNR = 10 \log_{10}\left(\frac{1}{MSE}\right) =  20\log_{10}N  - 10\log_{10} \Delta.
\end{equation*}

\begin{figure}
    \centering
    \begin{subfigure}[b]{0.09\linewidth}
        \fbox{\includegraphics[width=0.8\linewidth]{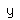}}
        \vspace{-5mm}
        \caption{{\scriptsize U+0079}}
    \end{subfigure}
    \begin{subfigure}[b]{0.09\linewidth}
        \fbox{\includegraphics[width=0.8\linewidth]{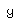}}
        \vspace{-5mm}
        \caption{{\scriptsize U+10e7}}
    \end{subfigure}
    \hspace{1mm}
    \begin{subfigure}[b]{0.09\linewidth}
        \fbox{\includegraphics[width=0.8\linewidth]{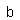}}
        \vspace{-5mm}
        \caption{{\scriptsize U+0062}}
    \end{subfigure}
    \begin{subfigure}[b]{0.09\linewidth}
        \fbox{\includegraphics[width=0.8\linewidth]{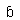}}
        \vspace{-5mm}
        \caption{{\scriptsize U+0253}}
    \end{subfigure}
    \hspace{1mm}    
    \begin{subfigure}[b]{0.09\linewidth}
        \fbox{\includegraphics[width=0.8\linewidth]{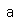}}
        \vspace{-5mm}
        \caption{{\scriptsize U+0061}}
    \end{subfigure}
    \begin{subfigure}[b]{0.09\linewidth}
        \fbox{\includegraphics[width=0.8\linewidth]{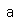}}
        \vspace{-5mm}
        \caption{{\scriptsize U+0430}}
    \end{subfigure}
    \\
    \begin{subfigure}[b]{0.09\linewidth}
        \fbox{\includegraphics[width=0.8\linewidth]{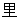}}
        \vspace{-5mm}
        \caption{{\scriptsize U+91cc}}
    \end{subfigure}
    \begin{subfigure}[b]{0.09\linewidth}
         \fbox{\includegraphics[width=0.8\linewidth]{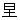}}
         \vspace{-5mm}
         \caption{{\scriptsize U+573c}}
     \end{subfigure}
    \hspace{1mm}
     \begin{subfigure}[b]{0.09\linewidth}
        \fbox{\includegraphics[width=0.8\linewidth]{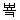}}
        \vspace{-5mm}
        \caption{{\scriptsize U+bfc8}}
    \end{subfigure}
    \begin{subfigure}[b]{0.09\linewidth}
         \fbox{\includegraphics[width=0.8\linewidth]{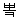}}
         \vspace{-5mm}
         \caption{{\scriptsize U+bf58}}
     \end{subfigure}
    \hspace{1mm}
     \begin{subfigure}[b]{0.09\linewidth}
        \fbox{\includegraphics[width=0.8\linewidth]{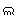}}
        \vspace{-5mm}
        \caption{{\scriptsize U+0b32}}
    \end{subfigure}
    \begin{subfigure}[b]{0.09\linewidth}
        \fbox{\includegraphics[width=0.8\linewidth]{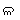}}
        \vspace{-5mm}
        \caption{{\scriptsize U+0b33}}
     \end{subfigure}     
     \caption{Examples of Unifont glyph images of Unicode characters. Top: Basic Latin characters and their homoglyphs. Bottom: CJK Unified Ideographs, Hangul syllables, and Oriya (Indo-Aryan language spoken in the Indian state of Odisha), and their homoglyphs.}
     \label{fig:image-ex}
\end{figure}

In the following, we show the processes we employed to construct the {\it SimChar} database.

\begin{figure*}[tbp]
\centering
 \begin{subfigure}[b]{0.05\linewidth}
 \centering
\fbox{\includegraphics[width=0.6\linewidth]{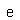}}
        \caption{$\Delta=0$}
\end{subfigure}
\begin{subfigure}[b]{0.05\linewidth}
 \centering
\fbox{\includegraphics[width=0.6\linewidth]{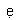}}
        \caption{$\Delta=1$}
\end{subfigure}
\begin{subfigure}[b]{0.10\linewidth}
 \centering
\fbox{\includegraphics[width=0.6\linewidth]{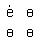}}
        \caption{$\Delta=2$}
\end{subfigure}
\begin{subfigure}[b]{0.15\linewidth}
 \centering
\fbox{\includegraphics[width=0.6\linewidth]{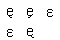}}
        \caption{$\Delta=3$}
\end{subfigure}
\begin{subfigure}[b]{0.25\linewidth}
 \centering
\fbox{\includegraphics[width=0.6\linewidth]{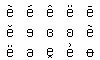}}
        \caption{$\Delta=4$}
\end{subfigure}
\begin{subfigure}[b]{0.10\linewidth}
 \centering
\fbox{\includegraphics[width=0.6\linewidth]{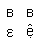}}
        \caption{$\Delta=5$}
\end{subfigure}
\begin{subfigure}[b]{0.25\linewidth}
 \centering
\fbox{\includegraphics[width=0.6\linewidth]{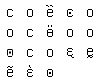}}
        \caption{$\Delta=6$}
\end{subfigure}
          \caption{Basic Latin lowercase letter `e' and characters under different values of the threshold $\Delta$. In this work, characters with $\Delta\leq 4$ are detected as homoglyphs.}
     \label{fig:delta-e}
\end{figure*}

\begin{figure}
    \centering
    \begin{subfigure}[b]{0.09\linewidth}
        \fbox{\includegraphics[width=0.8\linewidth]{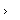}}
        \vspace{-5mm}
        \caption{{\tiny U+1be7}}
    \end{subfigure}
    \hspace{2mm}
    \begin{subfigure}[b]{0.09\linewidth}
        \fbox{\includegraphics[width=0.8\linewidth]{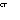}}
        \vspace{-5mm}
        \caption{{\tiny U+2df5}}
    \end{subfigure}
    \hspace{2mm}
    \begin{subfigure}[b]{0.09\linewidth}
        \fbox{\includegraphics[width=0.8\linewidth]{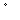}}
        \vspace{-5mm}
        \caption{{\tiny U+a953}}
    \end{subfigure}
    \hspace{2mm}
    \begin{subfigure}[b]{0.09\linewidth}
        \fbox{\includegraphics[width=0.8\linewidth]{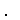}}
        \vspace{-5mm}
        \caption{{\tiny U+abec}}
    \end{subfigure}
     \caption{Examples of sparse characters.}
     \label{fig:image-sparse-ex}
\end{figure}

\begin{description}
\item[Step I] For the {52,457} characters in the intersection of the IDNA2008 draft and Unifont12, we represent the characters as bitmap images of $32 \times 32$ pixels, using the Unifont glyphs. Note that the original size of Unifont11 is $8 \times 8$ pixels for Latin characters and $16 \times 16$ for other characters.
Figure~\ref{fig:image-ex} presents the example of the generated Unifont glyph images where we intentionally chose visually similar pairs.

\item[Step II] For all the pairs in the pairwise combinations of the {52,457} characters, we compute the metric $\Delta$. 
If $\Delta$ is less than or equal to a threshold $\theta$, the two characters are identified as homoglyphs. In this work, we empirically derived a conservative threshold as $\theta = 4$; i.e., a pair of characters are detected as homoglyphs if $\Delta \leq 4$. Figure~\ref{fig:delta-e} shows examples of Unicode characters with various values of $\Delta$.
Although $\Delta=4$ would not indicate obvious {\it false positives} (i.e., those that should not be detected as homoglyphs), we can observe several {\it false negatives} (i.e., those that could be detected as homoglyphs) among characters with $\Delta \geq 5$. 
In Section~\ref{sec:evaluation}, we consider an evaluation of the validity of the threshold by presenting a human study.

\item[Step III] Finally, from the extracted pairs, we eliminate {\it sparse} characters that contain fewer than 10 black pixels. The threshold was empirically derived as a result of careful manual effort. In most cases, these characters are used for punctuation, spacing/nonspacing, or combining in various languages. Figure~\ref{fig:image-sparse-ex} presents examples of the eliminated characters.
\end{description}

After performing the four processes described above, we obtained a set of 12,636 characters. The set constitutes 13,126 pairs, which we named {\it SimChar}. 
As shown in Table~\ref{tab:charset_breakdown1}, the size of the intersection of {\it SimChar} and {UC} is fairly small, indicating that {\it SimChar} successfully adds new homoglyphs that have not been covered by {UC}. We also note that there are several characters that are not covered by {\it SimChar}, but are covered by {UC}. Thus, the two character sets can be used complementary to identify potential IDN homograph attacks.

\subsection{Characteristics of SimChar}

\begin{table}[tbp]
\caption{Number of homoglyphs of Latin letters (lowercase) contained in SimChar and UC $\cap$ IDNA.}
\label{tab:latinhomoglyphs}
\vspace{-4mm}
\begin{tabular}{c}
\begin{minipage}[b]{0.4\linewidth}
  \begin{tabular}{l|r|l|r|l|r} 
    \multicolumn{6}{c}{{\it SimChar}}\\
    \hline
     & \# &  & \# & & \#\\
    \hline
    `o' & 40  & `s' & 14 & `f' & 8\\
    `e' & 26  & `r' & 14 & `m' & 8\\
    `n' & 24  & `a' & 14 & `g' & 7\\
    `w' & 20  & `k' & 13 & `j' & 7\\
    `c' & 19  & `t' & 13 & `p' & 7\\
    `l' & 18  & `z' & 12 & `x' & 6\\
    `u' & 18  & `d' & 10 & `q' & 2\\
    `h' & 17  & `y' &  9 & `v' & 1\\
    `i' & 16  & `b' &  8 &     &  \\
    \hline\hline
    \multicolumn{3}{c}{Total}&  \multicolumn{3}{|c}{351}\\
  \hline
  \end{tabular}
\end{minipage}
\begin{minipage}[b]{0.1\linewidth}
\hspace{4mm}
\end{minipage}
\begin{minipage}[b]{0.4\linewidth}
  \begin{tabular}{l|r|l|r|l|r} 
    \multicolumn{6}{c}{{UC} $\cap$ {\it IDNA}}\\
    \hline
     & \# &  & \# & & \#\\
    \hline
    `o' & 34  & `c' &  4 & `p' & 3\\
    `l' & 12  & `d' &  4 & `x' & 3\\
    `y' & 10  & `g' &  4 & `j' & 2\\
    `i' &  9  & `f' &  4 & `n' & 2\\
    `u' &  9  & `a' &  3 & `z' & 2\\
    `w' &  8  & `b' &  3 &     & \\
    `v' &  6  & `e' &  3 &     & \\
    `s' &  5  & `h' &  3 &     & \\
    `r' &  5  & `q' &  3 &     &  \\
    \hline\hline
    \multicolumn{3}{c}{Total} & 
    \multicolumn{3}{|c}{141}\\
  \hline
\end{tabular} 
\end{minipage}
\end{tabular}
\end{table}

\noindent{\bf Homoglyphs of Latin Letters}
As the majority of popular websites make use of the 26 Latin letters to construct their primary domain names, it is essential to study the extent to which our homoglyph database covers the homoglyphs of Latin letters.
Table~\ref{tab:latinhomoglyphs} lists the results.
We first notice that {\it SimChar} successfully extracted new homoglyphs that have not been contained in UC. 
For instance, whereas the intersection of IDNA2008 and UC contains only three homoglyphs for the Basic Latin lowercase letter `e', {\it SimChar} contains 26 homoglyphs of `e' as shown in Figure~\ref{fig:delta-e}.
We also notice that several characters have many homoglyphs.
In total, SimChar contains 351 homoglyphs of Latin letters, whereas UC contains 141 of these homoglyphs.
In the {\it SimChar} dataset, the Basic Latin lowercase letter `o' has 40 characters that are visually similar to it, indicating that the character is ``vulnerable'' to an IDN homograph attack. 
We note that the intersection of the sets of homoglyphs for `o' for {\it SimChar} and {UC} contains {5} characters, implying that they cover different sets of homoglyphs of `o'; i.e., the majority of homoglyphs of `o' listed in {\it SimChar} were accented characters of `o', whereas the majority of homoglyphs of `o' listed in {UC} were characters of which the appearance resembles a circle.

\noindent{\bf Unicode Blocks}
In Unicode, a block is a contiguous range of code points. A block consists of hundreds to tens of thousands of characters. The characters contained in a block are typically associated with the writing systems in which the characters are used; e.g., the Basic Latin block consists of all the characters and control codes of the ASCII character set.
The majority of the blocks are classified into two planes: the Basic Multilingual Plane (BMP) and Supplementary Multilingual Plane (SMP).
In the BMP, the largest block is the CJK Unified Ideograph, the characters of which are used in the Chinese, Japanese, and Korean languages, and it contains more than 20 K of Chinese characters.

Table~\ref{tab:top5blocks} compares UC and {\it SimChar} with respect to their top-5 blocks.
Although the two scripts, CJK Unified Ideographs and Arabic are commonly found, the breakdown of these scripts differ from each other, indicating that the coverage of UC and {\it SimChar} is different. Our contribution is to automatically build {\it SimChar}, which can complement the manually compiled list of homoglyphs, i.e., {UC}. 
We note that the \url{.com} TLD is allowed to use characters from either of these blocks for IDN.

\begin{table}[tbp]
\caption{Top-5 Unicode blocks contained in {\it SimChar} (left) and UC $\cap$ IDNA (right).
CJK, CDM, Hangul, and CA are abbreviations of the CJK Unified Ideographs, Combining Diacritical Marks, Hangul Syllables, and Unified Canadian Aboriginal Syllabics, respectively.
}
\label{tab:top5blocks}
\centering
\begin{tabular}{c}
\begin{minipage}[b]{0.4\linewidth}
  \begin{tabular}{l|r} 
    \multicolumn{2}{c}{{\it SimChar}}\\
    \hline
    Block & \#chars \\\hline
    Hangul & 8,787\\
    CJK  & 395\\
    CA & 387\\
    Vai & 134\\
    Arabic & 107\\
  \hline
  \end{tabular} 
\end{minipage}

\begin{minipage}[b]{0.1\linewidth}
\hspace{4mm}
\end{minipage}

\begin{minipage}[b]{0.4\linewidth}
  \begin{tabular}{l|r} 
    \multicolumn{2}{c}{{UC} $\cap$ {\it IDNA}}\\
    \hline
    Block & \#chars \\\hline
    CJK  & 91\\
    CDM & 56\\
    Arabic & 52\\
    Cyrillic & 40\\
    Thai & 36\\
    \hline
   \end{tabular}
\end{minipage}
\end{tabular}
\end{table}

\section{Performance Evaluation}
\label{sec:evaluation}

This section presents our evaluation of the performance of the {\it ShamFinder} framework from the viewpoints of (1) human perception and (2) computational cost.

\subsection{Human Perception}

We evaluated the human perception of the homoglyphs listed in our {\it SimChar} database; i.e., to determine whether humans perceive their homoglyphs as confusing.
To this end, we employed a series of human study experiments using a crowd sourcing platform, Amazon Mechanical Turk (MTurk in short).
We designed two types of experiments. 
In our first experiment, we studied the effect of the threshold $\Delta$, which was introduced in Section~\ref{sec:simchar}, on the extent to which {\it SimChar} homoglyphs could be confused, i.e., their ``confusability.'' This experiment is intended to demonstrate the validity of the threshold we determined for detecting homoglyphs, i.e., $\Delta = 4$. 
Next, we compare the confusability of {\it SimChar} and UC, with the baseline of random pairs of characters.

\noindent{\bf Experimental Setup}
We created a crowd sourcing task that asks a participant whether pairs of two characters, which may contain homoglyphs, are confusing or distinct. 
Before performing the large-scale experiment, we carefully designed our experiment by conducting a series of pilot study trials that enabled us to adjust the wording of questions and answers. 
Several trials of the pilot study allowed us to obtain useful feedback from coworkers and participants, and we ultimately worded the question as ``{\it There are two characters shown in the image. Are they distinct or confusing?}.'' In terms of the answer, the following words were selected as the options for the five-level Likert scale score, ``1: {\it very distinct},'' ``2: {\it distinct},'' ``3: {\it neutral},'' ``4: {\it confusing},'' and ``5: {\it very confusing}.''
In this work, we refer to this score as the ``confusability score.''

\begin{figure}[tbp]
    \begin{center}
    \includegraphics[width=\linewidth]{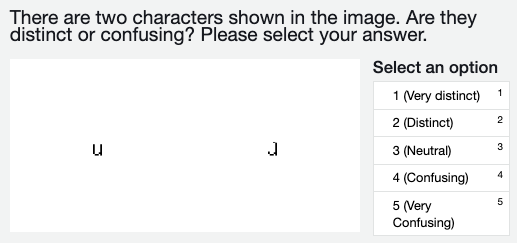}
    \caption{Screenshot of the MTurk task.}
    \label{fig:mturk-screenshot}
    \end{center}
\end{figure}

Figure~\ref{fig:mturk-screenshot} presents a screenshot of an assignment in the task presented to participants. The purpose of the assignment was to judge whether two characters contained in an image are distinct or confusing. 
Before conducting crowd sourcing experiments, we measured the average time to finish an assignment by ourselves and found an assignment to require approximately 15 seconds to complete, including the time to select an answer, submit it via the web interface, and wait for the page transition to the next assignment. On the basis of this observation, we set the reward per assignment as 0.05 USD, implying that the reward is equivalent to an average hourly compensation of 12 USD.
As the minimum wage in the USA is in the range of 7--12 USD / hour~\cite{uswage} (as of March 2019), we believe our payment configuration was appropriate, i.e., it was neither too low nor too high.

To ensure the quality of experiments, we used the following two criteria when recruiting participants: (1) the number of approved tasks of a participant should exceed 50 and (2) the participant should have a task approval rate greater than 97\%.

\begin{figure}[tbp]
    \begin{center}
    \includegraphics[width=\linewidth]{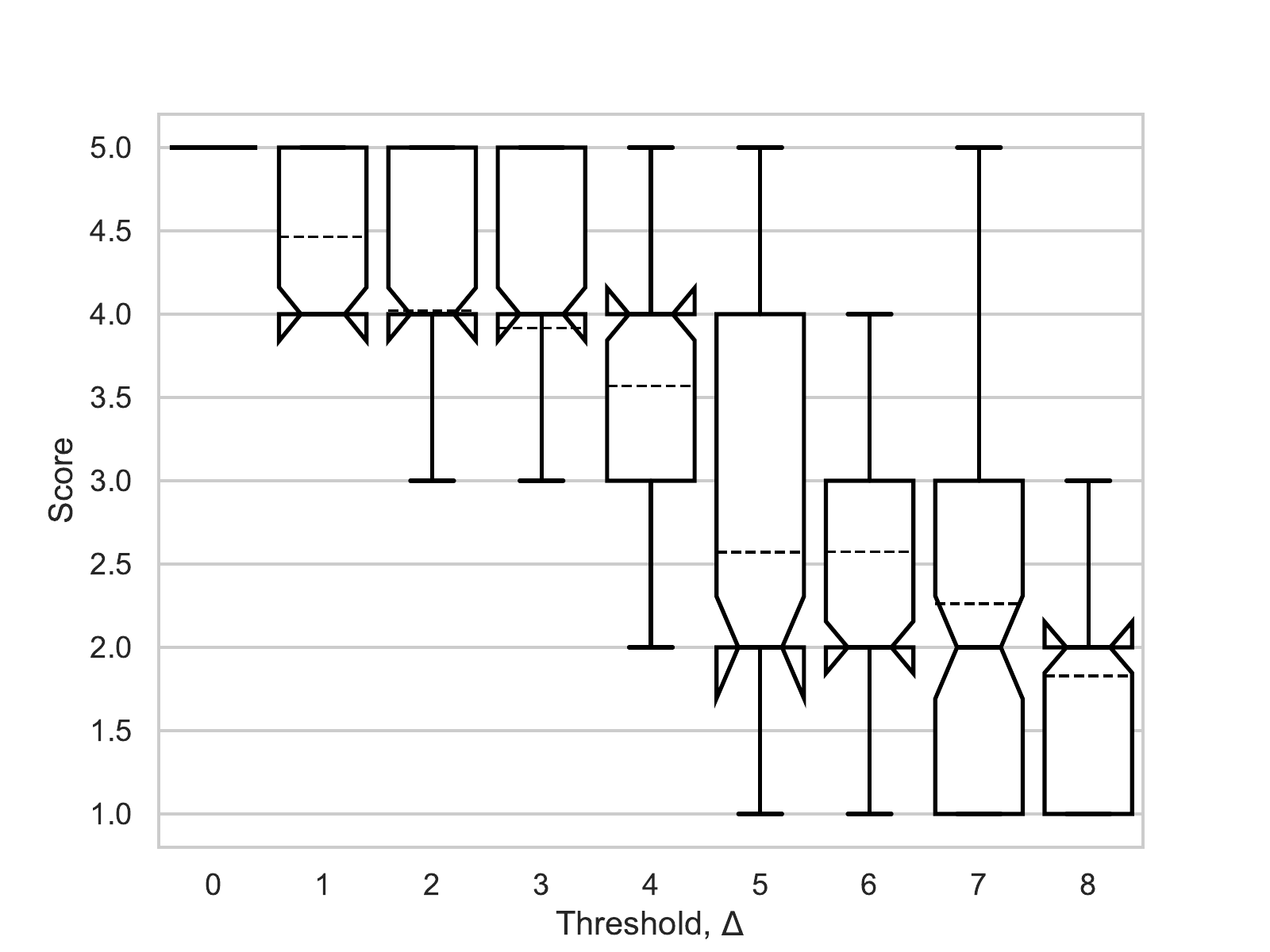}
    \caption{Boxplot of the confusability scores for pairs of SimChar homoglyphs for different values of the threshold, $\Delta$. The notch in each box represents the median and the dashed lines represent the mean values.
    Whiskers represent 1.5 IQR.}
    \label{fig:human-simchar}
    \end{center}
\end{figure}

To check whether a participant was careful when completing the task, we inserted dummy images that contain two completely distinct random characters. A participant who judged a dummy image as being either   
``4: {\it confusing}'' or ``5: {\it very confusing}'' had all their responses removed, assuming that the reliability of the participant was low. 
We likewise removed all the responses from participants who answered 
``1: {\it very distinct}'' or ``2: {\it distinct}'' to a homoglyph contained in {\it SimChar} with the threshold of $\Delta=0$, i.e., when the glyphs of the two characters were perfectly identical with the font we used (GNU Unifont).
Although this strategy may have aggressively removed the useful responses by a participant who accidentally made a single mistake, we decided to overcome the drawback by simply increasing the number of responses/samples.

\noindent{\bf Experiment 1: Threshold of SimChar}
We first studied the way in which the threshold, $\Delta$, affects human perception. 
In this experiment, we used homoglyphs of the Basic Latin letters (lowercase), the numbers of which are listed in Table~\ref{tab:latinhomoglyphs}.
For each letter, we extracted the glyphs with a distance of $\Delta\in \{0,\dots,8\}$. 
For each $\Delta$, we randomly sampled 20 pairs, where a pair consists of a letter and its potential homoglyph detected with the threshold $\Delta$. 
In addition, we added 30 of dummy pairs that contain two distinct letters randomly generated.
These $20 \times 9 = 180$ pairs of potential homoglyphs and 30 random pairs were judged by 10 participants (after the removal of unreliable participants).
In total, we obtained 900 effective responses for the 180 pairs.

Figure~\ref{fig:human-simchar} presents the result. 
As expected, the confusability score decreases as the threshold increases. 
When $\Delta=4$, the mean and median of the confusability score were 3.57 and 4, respectively. This observation implies that the homoglyphs detected with the threshold were mostly perceived as ``confusing.'' 
When $\Delta=5$, the mean and median of the confusability score were 2.57 and 2, respectively, implying that the detected homoglyphs were mostly perceived as ``distinct.'' 
On the basis of these observations, we adopted $\Delta=4$ as the threshold for extracting homoglyphs; i.e., glyphs with $\Delta \leq 4$ were detected as homoglyphs.
Although several pairs with the threshold of $\Delta = 5$ had a high confusability score, we adopted a conservative decision.
Extracting further confusable homoglyphs from these potential homoglyphs remains as a future task.

\begin{figure}[tbp]
    \centering
    \includegraphics[width=\linewidth]{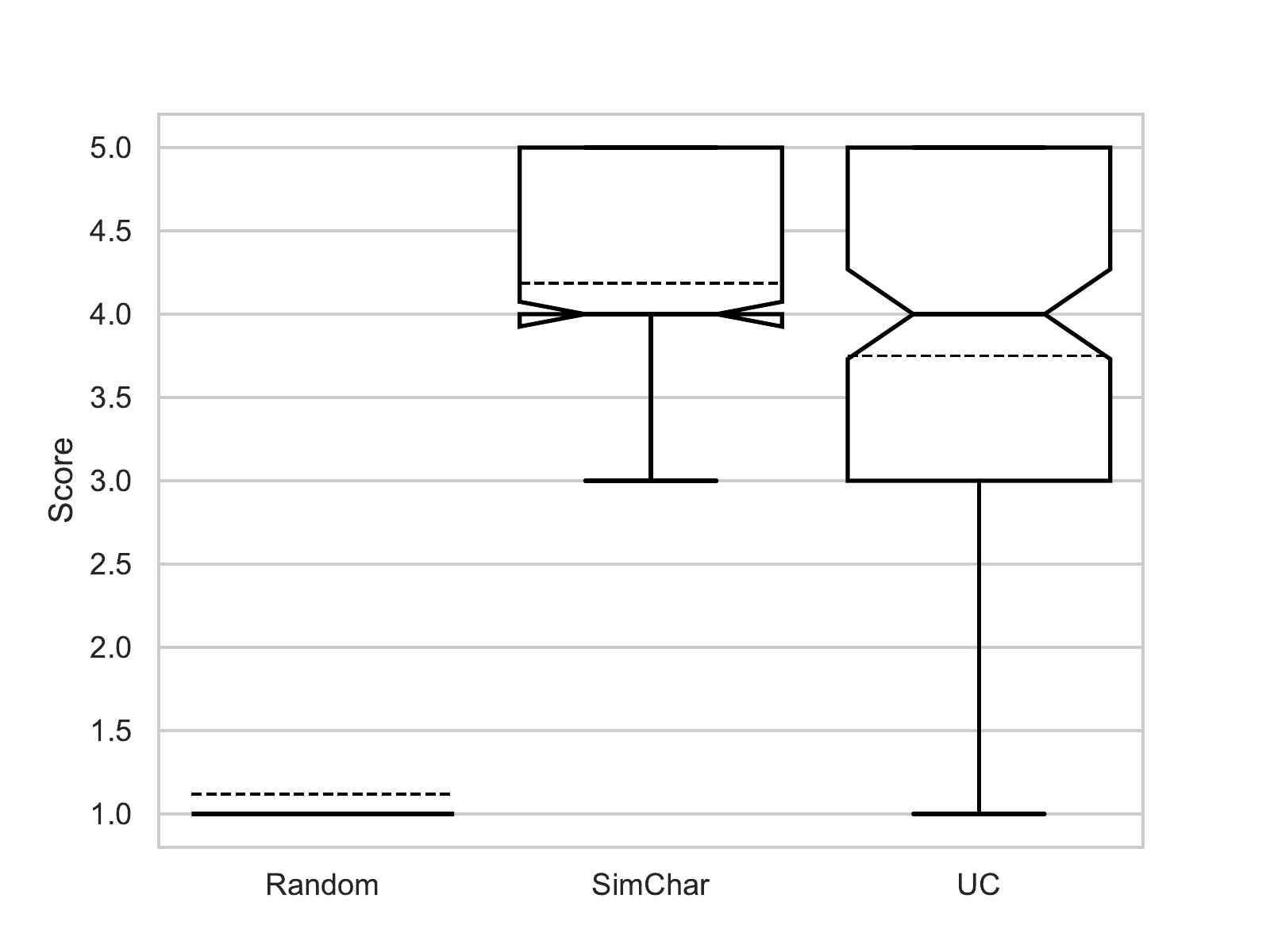}
    \caption{Boxplot of the confusability score for the pairs of three sets: Random (left), SimChar (middle), and UC (right). The threshold of SimChar was set to $\Delta\leq 4$. The boxplot configuration is the same as in Figure~\ref{fig:human-uc-simchar}.}
    \label{fig:human-uc-simchar}
\end{figure}

\noindent{\bf Experiment 2: Confusability of UC and SimChar.}
Next, we studied the confusability of UC in comparison with {\it SimChar} for which we repeated the same procedure shown above. 
We sampled 30 of the homoglyphs of the Basic Latin letters (lowercase) listed in UC.
These 30 pairs were judged by 28 participants (after the removal of unreliable participants).
In total, we obtained 513 effective responses for the 30 pairs sampled from UC. 
For {\it SimChar}, we compiled 486 effective responses for the $20 \times 5 = 100$ pairs of homoglyphs detected with $\Delta \leq 4$.

Figure~\ref{fig:human-uc-simchar} shows the result. 
For comparison, 513 of the effective responses for the 30 dummy pairs (Random) are also plotted. 
Although the confusable scores of the random pairs were mostly concentrated near the lowest option (``very distinct''), for both {\it SimChar} and UC, the median of the confusable score was 4, i.e., the homoglyphs of both databases were perceived as ``confusing'' on average. Note that the average confusable score for {\it SimChar} was larger than 4, whereas that for UC was smaller than 4, implying that the homoglyphs contained in {\it SimChar} were more confusable than those contained in UC.

\begin{figure}
    \centering
    \begin{subfigure}[b]{0.09\linewidth}
     \fbox{\includegraphics[width=0.8\linewidth]{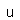}}
        \vspace{-5mm}
        \caption{{\tiny U+0075}}
    \end{subfigure}
   \hspace{1mm}
    \begin{subfigure}[b]{0.09\linewidth}
        \fbox{\includegraphics[width=0.8\linewidth]{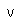}}
        \vspace{-5mm}
        \caption{{\tiny U+118D8}}
    \end{subfigure}
    \hspace{8mm}
    \begin{subfigure}[b]{0.09\linewidth}
        \fbox{\includegraphics[width=0.8\linewidth]{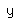}}
        \vspace{-5mm}
        \caption{{\tiny U+0079}}
    \end{subfigure}
    \hspace{1mm}
    \begin{subfigure}[b]{0.09\linewidth}
        \fbox{\includegraphics[width=0.8\linewidth]{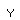}}
        \vspace{-5mm}
        \caption{{\tiny U+028F}}
    \end{subfigure}
    \hspace{8mm}     
   \begin{subfigure}[b]{0.09\linewidth}
     \fbox{\includegraphics[width=0.8\linewidth]{fig/U+0x79.png}}
        \vspace{-5mm}
        \caption{{\tiny U+0079}}
    \end{subfigure}
   \hspace{1mm}
    \begin{subfigure}[b]{0.09\linewidth}
        \fbox{\includegraphics[width=0.8\linewidth]{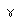}}
        \vspace{-5mm}
        \caption{{\tiny U+118DC}}
    \end{subfigure}
     \caption{Three pairs of homoglyphs listed in UC. They were most frequently judged as ``very distinct'' by the participants. They are homoglyphs of `u' (left),  `y' (middle), and `y' (right), respectively.}
     \label{fig:worstuc}
\end{figure}

Figure~\ref{fig:worstuc} presents three examples of UC pairs that attracted the lowest confusability score.
As these examples imply, several homoglyphs listed in UC have glyphs that could be perceived as distinct from the original letter, although some of the pairs could be semantically close. 
On the other hand, the homoglyphs listed in {\it SimChar} should have small differences by definition. 
These results led us to conclude that the homoglyphs listed in {\it SimChar} are actually perceived as confusable.

\subsection{Computation Cost of the ShamFinder Framework}

We first measured the time taken for constructing {\it SimChar}. Table~\ref{tab:time-simchar} summarizes the results.
As expected, the time for computing $\Delta$  for the pairwise combination of 52,457 characters, which is provided in Table~\ref{tab:charset_breakdown2}, was the most time-consuming step of the computation. 
For this computation, we used a multi-processing approach with the number of concurrent processes set to 15. We used an off-the-shelf server with an Intel Xeon CPU E5-2620 v2 (2.10 GHz) and 62 GB memory.
In practice, we would need to update {\it SimChar} when the Unicode standard adds a new set of glyphs or we incorporate a new set of fonts to be analyzed. That is, the frequency of updating {\it SimChar} should be reasonably low; e.g., Unicode version 12.0 was released one year after the release of version 11.0. The new version added 553 characters to those in the previous one.

Next, we measured the time to extract IDN homographs using the {\it ShamFinder} framework. 
To extract IDN homographs of the Alexa top-10k domains from the 141 M of {\tt .com} TLD domain names (see Table~\ref{tab:zonefile} for reference) required 743.6 seconds, i.e., on average, each reference domain name was inspected in $0.07 (= 743.6/10,000)$ seconds, which is sufficiently fast to block a suspicious, newly found IDN homograph attack in real time.

\begin{table}[tbp]
\caption{Time taken for constructing SimChar.}
\label{tab:time-simchar}
\centering
  \begin{tabular}{l|r} \hline
  Process                           &  Time  \\\hline\hline
  Generating images                 & 79.2 seconds \\
  Computing $\Delta$ for all the pairs  &  10.9 hours \\ 
  Eliminating sparse characters     &  18.0 seconds \\ 
  \hline
  \end{tabular} 
\end{table}

\section{Data Sources}
\label{sec:data}
In this section, we describe the data sources used for our analysis of IDN homographs.

\subsection{Reference Domain Names}

The aim of an IDN homograph attack is to attract a victim to a malicious website by using a homograph that is visually identical to the domain name of a legitimate website. As such, the natural assumption is that an attacker creates an IDN homograph of a domain name used for a popular website.
In fact, other deception techniques such as ``typosquatting'' or ``brandjacking'' also target widely recognized domain names~\cite{DBLP:conf/uss/SzurdiKCSFK14,DBLP:conf/ndss/AgtenJPN15}. 
As a reference of well-known popular domain names, we adopted Alexa Top Sites~\cite{alexa}; 
namely, we extracted the top-10K of .com domains from the Alexa ranking list.

\subsection{Extracting IDNs}

Although many domain name spaces are available in the Internet, in this study, we focused on domain names under the {\tt .com} TLD for the following three reasons.
First, the majority of popular websites are attributed to this TLD. As the word ``dot-com bubble'' symbolizes, {\tt .com} has become the most popular TLD since the early 2000s.
Although .com was originally intended for commercial usage, it eventually became available for general purposes.
Second, as shown below, the majority of malicious IDNs are also attributed to this TLD. 
Finally, as {\tt .com} TLD is globally popular, it permits a large number of Unicode blocks to be used for IDNs.
According to IANA's IDN tables~\cite{idn-table}, under the .{\tt com} TLD, characters across 97 different Unicode blocks can be used for IDNs as of May 2019. This fact implies that for {\tt .com} TLD, an attacker can register an IDN homograph that contains homoglyphs sampled from various Unicode blocks.

To search for IDN homographs, we first needed to extract registered IDNs. 
To this end, we used the DNS zone file maintained by the registries of the {\tt .com} TLD --- Verisign~\cite{verisign}. 
The DNS zone file lists all the registered domain names with their NS records. 
We complemented the zone file by using another list of domain names named \url{domainlists.io}~\cite{domainlists-io}. 
The union set of the two lists contains 141.2 M of unique domain names. 
As mentioned above, we can extract IDNs by searching for domain names starting with the prefix ``\verb|xn--|''.


\begin{table}[tbp]
\caption{Summary of domain name lists and the number of IDNs they contained.}
\label{tab:zonefile}
\tabcolsep=1mm
\centering
\begin{tabular}{l||r|r|r} \hline
  Data  &  Number of    & Number of  & Collection\\ 
         &  domain names & IDNs       & time\\\hline\hline
  zone file~\cite{verisign} & 140,900,279   &  952,352 (0.67\%)  &  May 2019\\
  {\tt domainlists.io}~\cite{domainlists-io} & 139,667,014  &  953,209 (0.73\%)  &  May 2019\\
  \hline\hline    
  Total (union) & 141,212,035  &  955,512 (0.67\%)  &  --\\
  \hline
\end{tabular} 
\end{table}

Table~\ref{tab:zonefile} summarizes the number of domain names/IDNs for each dataset. We first notice that a non-negligible number of IDNs are currently registered in the .com TLD, implying the widespread adoption of IDN in the wild.
Next, we examined the languages used in those IDNs to understand which Unicode blocks are widely used in the IDNs.
To detect the language used in a string, we leveraged a tool known as LangID~\cite{langid}, which is a Python module that can detect the most plausible language out of 97 distinct languages for a given string. 
Table~\ref{tab:lang-detect} presents the results.
We see that east Asian languages (Chinese, Japanese, and Korean) are dominantly used for composing IDNs wheres several European languages are also popular for this purpose. 
This observation implies that the demand for the use of native languages is ubiquitous.

 \begin{table}[tbp]
     \caption{Top languages used for IDNs.}
     \label{tab:lang-detect}
     \centering
     \begin{tabular}{c|lrr}
     \hline
     Rank & Language & Number   & Fraction (\%) \\
     \hline\hline
     1 & Chinese  & 443,865 &  46.5 \\
     2 & Korean   & 101,711 &  10.6 \\
     3 & Japanese &  88,970 &  9.3 \\
     4 & Germany  &  53,776 &  5.6\\
     5 & Turkish  &  35,288	&  3.6 \\
     \hline
     \end{tabular}
 \end{table}
 
\section{Detecting IDN Homographs with the ShamFinder Framework.}
\label{sec:measurement}

In this section, we apply the {\it ShamFinder} framework to the data we described in the previous sections.
We first studied the IDN homographs that targeted popular domain names that reside in the .com TLD. 
We then studied the malicious IDN homographs detected by our approach. 
In comparison to the existing approach, we compared the number of detected malicious IDN homographs by changing the homoglyph database. As discussed in Section~\ref{sec:related}, the previous approach to detecting IDN homographs proposed by 
Quinkert et al.~\cite{Quinkert19} leveraged {UC} as their homoglyph database. That is, we can directly compare the IDN homograph detection performance between their approach ({UC} only) and ours ({UC} and {\it SimChar}).

\subsection{Statistics of the IDN Homographs}

\begin{table}[tbp]
    \caption{Number of detected IDN homographs for ASCII domains.}
    \label{tab:idn-ascii}
     \centering
     \begin{tabular}{l|r}
     \hline
     Homoglyph DB & Number \\
     \hline\hline
     {UC}  &  436\\
     {\it SimChar} & 3,110\\
     {UC} $\cup$ {\it SimChar} & 3,280\\
     \hline
    \end{tabular}
\end{table}

Table~\ref{tab:idn-ascii} presents the number of detected IDN homographs targeting ASCII-character domain names.
When we used {UC}, the {\it ShamFinder} framework detected 436 IDN homographs out of the 955 K IDNs registered in the .com TLD. 
On the other hand, when we used {\it SimChar}, more than 3,110 of IDN homographs were detected. 
In total, by using both homoglyph databases, we detected 3,280 IDN homographs, which is approximately eight times more than those detected with {UC}.
Thus, the adoption of {\it SimChar} as the homoglyph database enables us to detect more IDN homographs than existing approaches such as that of Quinkert~\cite{Quinkert19}.

Table~\ref{tab:top5domain} presents the the top-5 domain names that have the most IDN homographs. 
Three of these domains, google.com, amazon.com, and facebook.com are all popular domains; however, the two other domains, myetherwallet.com and allstate.com are not that popular compared to the other three domains. In fact, the first three domains are ranked among the top-10 domains in the Alexa ranking, whereas the other two domains are ranked 7,400th and 5,148th among the .com TLD domains in the Alexa ranking, respectively. 
This observation demonstrates that IDN homograph attacks not only target very popular websites, they also target other moderately popular websites, implying that starting with a small list of reference domains may not be effective for IDN homographs that target minor domains. 
We discuss this issue below (Section~\ref{sec:revert}).

In the following, we analyze the IDNs that are currently active. First, we checked the NS records for the 3,280 homograph IDNs we detected. We found 2,294 domain names with NS records, while other domain names did not have NS records due to some reasons such as expiration, non-registration, etc. Of the 2,294 domain names, 385 domain names did not have A records. For the remaining 1,909 domain names, we performed port scans to the ports TCP/80 and TCP/443. Table~\ref{tab:portscan} shows the results. We found that the 1,647 IDN homographs we detected were reachable through the HTTP or HTTPS; i.e., roughly half of the detected IDN homographs were active.

 \begin{table}[tbp]
     \caption{Top-5 ASCII domain names that have the most IDN homographs.}
     \label{tab:top5domain}
     \centering
     \begin{tabular}{c|lr}
     \hline
     Rank & Domain name & \# homographs\\
     \hline\hline
     1 & {\tt myetherwallet.com} & 170  \\
     2 & {\tt google.com }       & 114  \\
     3 & {\tt amazon.com }       & 75   \\
     4 & {\tt facebook.com }     & 72   \\
     5 & {\tt allstate.com }     & 68   \\
     \hline
     \end{tabular}
 \end{table}
 
  \begin{table}[tbp]
     \caption{Port scan results for the detected IDN homographs.}
     \label{tab:portscan}
     \centering
     \begin{tabular}{l|r}
     \hline
    Ports  & \# domain names\\
    \hline\hline
    TCP/80  & 1,642   \\
    TCP/443  &  700   \\
    TCP/80 \& TCP/443   & 695    \\
    \hline        
    Total (unique)   & 1,647 \\
    \hline
 \end{tabular}
 \end{table}

\subsection{Deep Inspection of the Active IDN Homographs}
In this section, we further inspect the characteristics of the active IDN homographs we found in the previous subsection.
In the following, we show the analyses from two aspects: (1) analysis of the popular IDN homographs and (2) classification of IDN homographs.

\vspace{-2mm}
\quad\\
\noindent{\bf (1) Analysis of the popular IDN homographs}\\
~To study how the active IDN homographs have been accessed by end users, we focus on the ``popular'' IDN homographs that likely attracted large number of end-users.
To this end, we performed the analysis using the passive DNS~\cite{passivedns}, which is a DNS monitoring system that is composed of several working DNS cache servers. A passive DNS system provides useful statistics such as the number of cumulative name resolutions for each domain name.
We note that the statistics provided by a passive DNS system reflects sampled data collected at a set of cache servers contributing to the system. Therefore, the actual numbers of DNS lookups over the entire Internet should be much larger than those obtained from a passive DNS system.
We also note that the number of web accesses and number of DNS resolutions are different. However, we believe that the number of DNS resolutions is correlated with the popularity of a domain name,
given that every first web query should be preceded by a DNS query.

Table~\ref{tab:top10domains} shows the top-10 domain names that had the largest numbers of DNS lookups. 
We studied the categories of the websites running on the IDNs by manual inspection. 
We found that of the top-10 IDNs, four of them targeted {\tt gmail.com}.
In particular, the top IDN, {\tt gmaıl[.]com} was an active phishing site and had a large number of name resolutions, implying that there have been a large number of end-users who accessed the phishing website\footnote{As of September 2019, this website was still in operation. We have reported about the website to the security vendors.}.
We found that the website under the IDN employed a cloaking technique to redirect a visitor to the different websites according to the User-agent of the visitor's browser.
We also found that the majority of the IDNs were parked domains; these were used for monetizing through advertisements and/or were reserved for resale.

In Table~\ref{tab:top10domains}, the columns ``MX,'' ``Web link,'' and ``SNS'' represent, 
whether there was a generic website linking to the IDN homograph, and whether there was a web link pointing to the IDN homograph on popular SNS websites such as Twitter. We used the search engines for the latter two analyses. We found that the IDN homographs that target domain names used for email services such as {\tt gmail.com} and {\tt yahoo.com} have MX records either in the past or in the present. We also saw that several IDN homographs have appeared in public webspace, including SNS. These observations imply that the owners of these IDN homographs have attempted to make the IDN homographs publicly visible.

\begin{table}[tbp]
     \caption{Top-10 active IDNs that had the largest numbers of DNS resolutions. \CIRCLE indicates that there is an active MX record registered. \LEFTcircle indicates that there was a MX record in the past.}
     \label{tab:top10domains}
     \tabcolsep=1mm
     \centering
     \begin{tabular}{l|l|r|c|c|c}
     \hline
     Domain name & Category  & \#resolutions & MX & Web link & SNS\\
     \hline\hline
    {\tt gmaıl[.]com}  & Phishing & 615,447  & \LEFTcircle & & \\
    {\tt döviz[.]com}  & Portal   & 127,417  &  & \checked & \checked\\
    {\tt ģmail[.]com}       & Parked  &  74,699 & \LEFTcircle& & \\
    {\tt gmàil[.]com}       & Parked  &  63,233 & \CIRCLE  & & \\
    {\tt expansión[.]com}   & Parked  &  56,918 & \LEFTcircle &\checked & \checked\\
    {\tt gmaiĺ[.]com}       & Parked  &  49,248 & \CIRCLE  & & \\
    {\tt yàhoo[.]com}        & Parked  & 44,368 & \LEFTcircle & & \\
    {\tt shädbase[.]com}     & Parked  & 38,556 & & \checked & \checked \\
    {\tt youtubê[.]com}      & Sale    & 37,713 & \CIRCLE  & & \\
    {\tt perú[.]com}         & Parked  & 36,405 & \CIRCLE & & \checked \\
    \hline
   \end{tabular}
 \end{table}

\vspace{-2mm}
\quad\\
\noindent{\bf (2) Classification of IDN homographs}\\

~We now attempt to classify the 1,647 active IDN homographs that responded to either TCP/80 or TCP/443.
To this end, we make use of a list of NS records for the domain parking companies, screenshots of the websites, and VirusTotal~\cite{virustotal}, which is an online virus scanner. 
To compile a list NS records for the domain parking companies, we leverage the list and methods proposed in~\cite{ndss2015,domainchroma}.
We added several NS records and ended up 17 of NS records used for domain parking.

Next, for the remaining IDN homographs that were not attributed to domain parking, we accessed to the corresponding websites via the two schemes, HTTP and HTTPS, and took screenshots using the puppeteer~\cite{puppeteer}, which is a headless browser that provides APIs to control Chrome or Chromium.  
Based on the characteristics of screenshots and HTTP responses, we classified the websites
into the following five categories: ``For sale,'' ``Redirect,'', ``Normal,'', ``Empty,'' and ``Error,'' which represent 
a website that encourages you to buy the domain, 
a website that redirects to another website, 
a website that displays something legitimate successfully, 
a website that displays nothing, 
and a website that failed to get a screenshot due to a timeout or other reasons, respectively.

Table~\ref{tab:homograph_class} shows the results. 
We found that 693 (42\%) of the websites running on IDN homographs were used for business (``Domain parking'' or ``For sale'').
We also found that 338 (21\%) of the websites running on IDN homographs were redirected to other websites having different domain names.
We further analyzed these 338 websites using VirusTotal and manual inspection of the screenshots.
Table~\ref{tab:homograph_redirect} shows the breakdown of the websites with redirect.
Brand protection indicates that a website running on a homograph domain name is redirected to the website running on the corresponding original domain name. That is, the owner of the original domain name has registered the homograph to protect their brand. We found that while the majority of the redirected domain names were attributed to either brand protection or legitimate websites, we found 35 of them were detected as malicious websites.

\begin{table}[tbp]
    \caption{Classification of the active IDN homographs.}
    \label{tab:homograph_class}
    \centering
    \begin{tabular}{l|r}
    \hline
    Category  & Numbers \\
    \hline\hline
    Domain parking & 348 \\
    For sale     & 345 \\
    Redirect     & 338  \\
    Normal        & 281 \\
    Empty        & 222 \\
    Error        & 113  \\\hline\hline
    Total        & 1,647 \\
    \hline
   \end{tabular}
\end{table}

\begin{table}[tbp]
    \caption{Classification of the active IDN homographs.}
    \label{tab:homograph_redirect}
    \centering
    \begin{tabular}{l|r}
    \hline
    Category  & Numbers \\
    \hline\hline    
    Brand protection    & 178\\
    Legitimate website &  125\\
    Malicious website  & 35  \\
    \hline\hline
    Total              & 338  \\
    \hline
   \end{tabular}
\end{table}

\subsection{Malicious IDN Homographs}

To check whether the detected IDN homographs have been used for malicious purposes, we leveraged three different sources of blacklists, hpHosts~\cite{hphosts}, Google Safe Browsing (GSB)~\cite{gsb}, and Symantec DeepSight~\cite{symantec-deepsight}; of the three lists, hpHosts, which is a community-based database, had the largest number of entries as we collected data spanning several years. As GSB and Symantec DeepSight are databases maintained by commercial companies, they provide lists of malicious domains that have been inspected by security experts with high confidence. 
We applied the blacklists to 3,280 of detected IDN homographs, which include non-active domains.
Table~\ref{tab:maliciousIDN} lists the results. 
We note that the numbers shown in the table do not include ones shown in the previous subsection; the previously found malicious websites had redirected URLs.
By incorporating {\it SimChar} into the homoglyph DB, the number of detected malicious IDN homographs increased.

 \begin{table}[tbp]
     \caption{Number of malicious IDN homographs.}
     \label{tab:maliciousIDN}
     \centering
     \begin{tabular}{l|r|r|r}
     \hline
     Homoglyph DB & hpHosts & GSB & Symantec  \\
     \hline\hline
     {UC}      &    28    &   2  &      1      \\
     {\it SimChar} &     222   &  12   &    7        \\
     {UC} $\cup$ {\it SimChar} & 242        &  13    &  8          \\
     \hline
     \end{tabular}
 \end{table}

\subsection{Reverting to Original Domains}
\label{sec:revert}

Although we begin with a reference domain name list to search for IDN homographs, this approach may not detect IDN homographs if a non-popular website is targeted. 
Therefore, if we find a malicious domain name, which is composed as an IDN, it is useful to be able to identify the original domain name targeted by the IDN homograph attack. Otherwise, we cannot trace the possible damage caused by the attack.
Thanks to the homoglyph database we developed, we can revert to the possible original domain name by replacing a homoglyph with the corresponding Basic Latin letter.
We reverted the malicious IDNs to the original domain names and removed those were contained in the Alexa top-1k domains. We ended up 91 of malicious IDNs whose original domains were {\it not} contained in the Alexa list. This observation indicates that there were non-negligible number of malicious IDN that targeted non-popular websites. Our approach can automatically revert such domains.

\section{Discussion}
\label{sec:discussion}

In this section, we first discuss the limitations of our work, after which 
we consider effective countermeasures against the threat of an IDN homograph attack.

\subsection{Limitations}
The primary contribution of this study was to build an automated framework that can detect a Unicode homoglyph and an IDN homograph. 
Below we discuss several limitations of the approaches we followed for evaluating our framework as well as their future extensions.

\noindent{\bf Confusability Test} 
In this work, we evaluated the confusability of homoglyphs by a single character, i.e., participants judged whether a potential homoglyph is confusable or distinct by viewing a pair of characters. However, as homoglyphs are generally abused in a word or even in a sentence, we may also need to study the confusability of homoglyphs by using words or sentences because this context may affect the user's perception. 
The context-aware evaluation of the confusability of a homoglyph is left for future study.

\noindent{\bf Font Type} 
In this work, we leveraged GNU Unifont, which is a bitmap-based font. GNU Unifont is one of the widely available Unicode glyphs with a wide range of coverage, but many other Unicode fonts are available in the wild, e.g., Noto font~\cite{notofont}, which is a scalable font.
As our framework is automated, it would be straightforward to extend our evaluation to other font families. 
This would be a future task.

\noindent{\bf Measurement Target} 
Our measurement study focused on the world's most popular TLD, {\tt .com}, yet many other TLDs are used in the wild.
For instance, the blacklists we used in this work contain 1,054 of domain names attributed to the 
`рф' TLD, which is the Cyrillic country code TLD for the Russian Federation. 
Studying such class of malicious IDNs from the viewpoints visual deception is left for future study.
In addition, although current IDN homograph attacks are mainly targeted at ASCII domains, IDNs that contain non-ASCII characters are emerging. Such IDNs may contain ideographs such as Hieroglyphs. 
Our approach can cover homoglyphs consisting of any characters including the ideographs. 
Studying these potential targets of homograph attacks and their threats would also be a future topic.

\subsection{Countermeasures against IDN Homograph Attacks}

As we have shown in Section~\ref{sec:background}, countermeasures against an IDN homograph attack implemented in modern browsers have the following drawbacks: 
if an IDN violates a rule of permitted characters, the countermeasure forcibly represents the IDN in the form of Punycode, which is not a user-friendly expression. This countermeasure may not provide a user with any indication of the {\it context} behind such transcoding. 
Moreover, a countermeasure is not effective against non-IDN homographs where homoglyphs reside in the same Unicode block; i.e, the IDN conforms with the rule of permitted characters. 

\begin{figure}[tbp]
\begin{center}
\includegraphics[width=\linewidth]{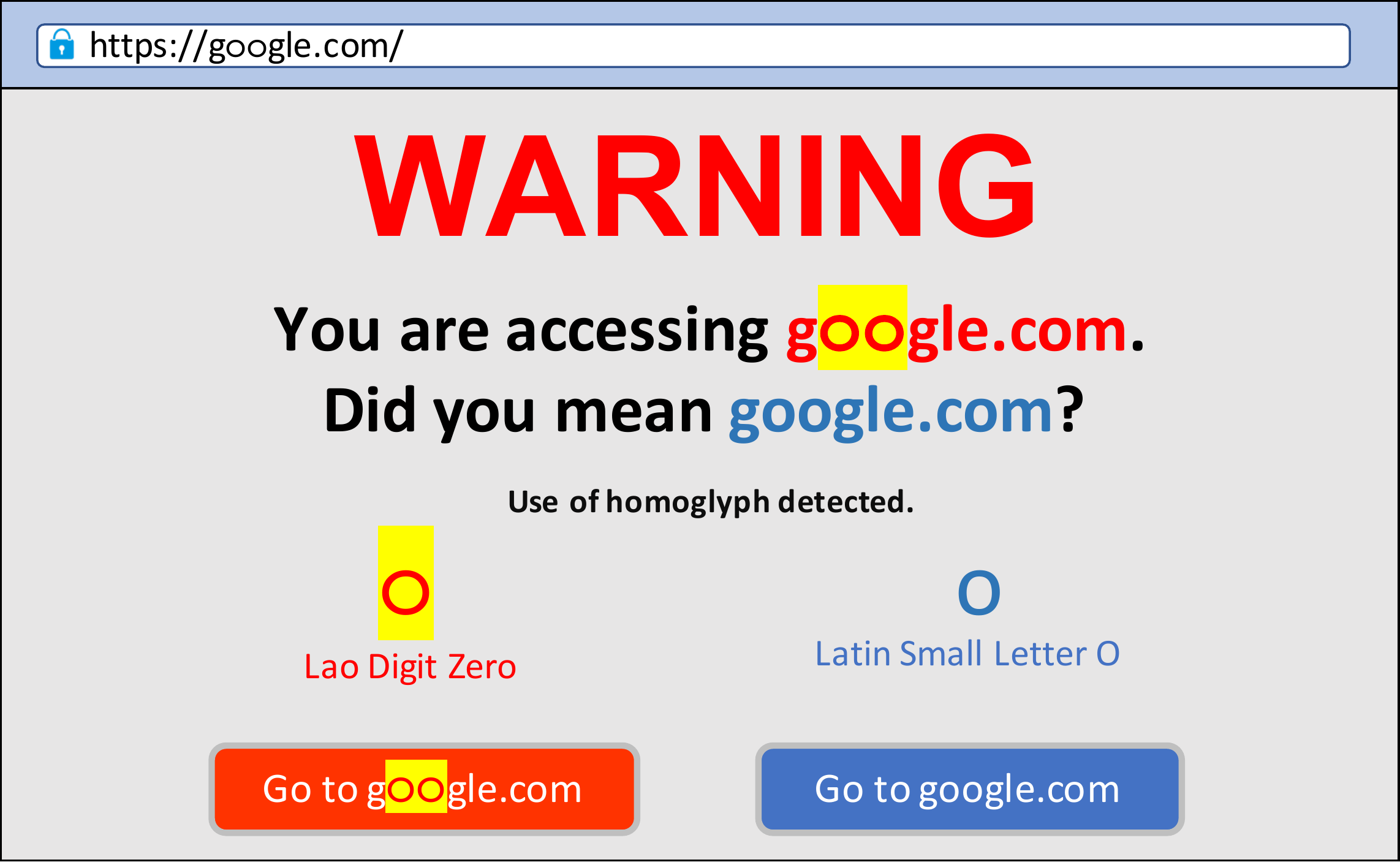}
\caption{Image of the UI presenting countermeasures against an IDN homograph attack based on the homoglyph database.}
\label{fig:poc}
\end{center}
\end{figure}

To explicitly inform the user of the possibility of an IDN homograph attack with a reasonable context would require the user to be presented with the Unicode representation, instead of forcibly converting the IDN to Punycode. To this end, we could adopt a user interface (UI) that emphasizes the difference between the original domain name and the potential IDN homograph. 
Figure~\ref{fig:poc} presents an image of such a UI, which could be implemented with the aid of homoglyph databases such as {\it SimChar} and {UC}.
We note that sizes of SimChar and UC are small enough to be embedded into a client program such as Browser extension/plug-in.. 
This UI would enable a user to understand which part of a domain name is replaced by which character. This information would be expected to play a vital role in informing the user about the possible threat of a phishing attack caused by an IDN homograph. More importantly, as an IDN is designed to provide a user-friendly expression of a domain name by using native languages, forcibly converting an IDN to Punycode would significantly impair the user experience. 
We expect the adoption of such an interface to improve users' awareness of the possible threats posed by an IDN homograph attack; i.e., they would become more knowledgeable regarding the context of the presented domain names and be more aware of possible threats.  
Implementation and evaluation of such a method could be the subject of further study.

\subsection{Ethical Considerations}
In Section~\ref{sec:evaluation}, we performed human study to assess the human perception on the detected homoglyphs.
Before conducting our human study experiments, we carefully followed the checklist provided by our institutional IRB and concluded that our experiments conformed with the principles of the research ethics. 
The fact that our user study does not collect any personally identifiable information nor privacy-sensitive information also justifies our conclusion. We also cared the amount of reward for the participants, considering the time to complete a task and minimum wage. 

\section{Related Work}
\label{sec:related}

In this section, we discuss related work in terms of IDN homograph detection methods and their measurement studies. 

\subsection{IDN Homograph Detection}

Several studies have led to the proposal of methods to detect IDN homographs. 
The approaches they followed are broadly classified into two types: image-based and character-based. 

\noindent{\bf Image-based IDN Homograph Detection}
As an IDN homograph exploits the visual similarity between characters, it is natural to apply image-based analysis for detecting these homographs. 
Liu et al.~\cite{DBLP:conf/dsn/LiuLLLDHZ18} generated images corresponding to 1.4 million registered IDNs and reference domain names extracted from the top 1,000 domain names listed on Alexa Top Sites. They then detected 1,516 IDN homographs based on the visual similarities between images. Furthermore, they found an additional 42,671 IDNs that were visually similar to the reference domain names but were still unregistered. Unfortunately, details of their detection methods and settings are not provided in their paper.
Sawabe et al.~\cite{sawabe18} developed a method to detect IDN homographs by leveraging optical character recognition (OCR). The method replaced non-ASCII characters in IDNs with similar ASCII characters using OCR-based image recognition and detected IDN homographs if the replaced IDNs corresponded with reference domain names on Alexa Top Sites.

\noindent{\bf Character-based IDN Homograph Detection}
A few researchers adopted the character-based approach. 
To the best of our knowledge, only two previous studies~\cite{Quinkert19,DBLP:conf/imc/TianJ0Y018} attempted to apply this approach. 
Quinkert et al.~\cite{Quinkert19} searched IDN homographs based on a list of homograph pairs, which is equivalent to the homoglyph DB using {UC} in our study, and detected 2,984 IDN homographs targeting 810 reference domain names. 
Tian et al.~\cite{DBLP:conf/imc/TianJ0Y018} developed a detection method based on {UC} to identify IDN homographs.
As shown in Section~\ref{sec:evaluation}, our homoglyph DB, \textit{SimChar} outperformed {UC}-based detection in the sense that the homoglyphs of {\it SimChar} were perceived to be more confusing than those of {UC} while maintaining high coverage of homoglyphs; thus, our method complements previous work to cover IDN homographs more comprehensively.

\subsection{Measurement Study of IDN Homograph Attacks}
Apart from the IDN homograph detection method described above, several researchers have performed measurement studies of IDN homograph attacks in the wild. 
In 2006, Holgers et al.~\cite{DBLP:conf/usenix/HolgersWG06} conducted a passive measurement study on a campus network to search for IDN homographs accessed by users. They also used active DNS probing to detect registered IDN homographs for a limited number of reference domains. 
Tian et al.~\cite{DBLP:conf/imc/TianJ0Y018} studied domains created by various types of domain squatting techniques including IDN homographs to detect phishing websites that exploit homographs in the wild. 
Le Pochat et al.~\cite{DBLP:conf/pam/PochatGJ19} defined the concept of IDNs that owners of brands with diacritical marks would like to use and generated 15,276 such IDNs. They found that 43\% of them were available for registration in 2019. 
Chiba et al.~\cite{chiba2019domainscouter} performed a measurement study to demonstrate that there are many IDN homograph attacks targeting non-English brands or combining other domain squatting methods.

These previous studies mainly focused on the measurement of IDN homographs. We believe our character-based approach to comprehensively detect IDN homographs could be readily applied to these studies, and thus could complement them to provide a more comprehensive understanding of IDN homographs. 

\section{Conclusion}
\label{sec:conclusion}

This work led to the development of a new framework named {\it ShamFinder}, which is useful for detecting IDN homographs efficiently. 
The key technical contribution of our work was the construction of a new homoglyph database named {\it SimChar}, which can be updated without requiring time-consuming manual efforts.
As {\it SimChar} is portable, it can be implemented in various systems/platforms as a key component of countermeasures against the threat of IDN homograph attacks. 
Noteworthy is that {\it SimChar} could be used for other promising security applications such as detecting obfuscated plagiarism, which exploits Unicode homoglyphs.  
We release the code and data of {\it ShamFinder}~\cite{shamfinder}.
Our future work includes the extension of our study; i.e., extending the domain name space to be explored, extending the font sets, studying the confusability of non-ASCII homoglyphs, etc.


\bibliographystyle{plain}
\bibliography{ref}

\end{document}